\documentstyle[12pt]{article}
\date{November 17, 1996}

\title{Variation in polarization of high-energy $\gamma$-quanta
traversing a bunch of polarized laser photons}

\author{G.L.~Kotkin\thanks{Electronic address:
kotkin@phys.nsu.nsk.ru}
$\;$ and V.G.~Serbo\thanks{Electronic address: serbo@math.nsc.ru}\\
{\it Novosibirsk State University, 630090, Novosibirsk, Russia}}

\begin{document}
\maketitle

\begin{abstract}
The elastic light-light scattering below the threshold of the
$e^+e^-$ pair production leads to a variation in polarization of
hard $\gamma$-quanta traversing without loss a region where the
laser light is focused. Equations are obtained which determine
the variation of Stokes parameters of $\gamma$-quanta in this
case, and their solutions are given. It is pointed out that this
effect can be observed in the experiment E-144 at SLAC. It should
be taken into account (and, perhaps, it can be used) in
experiments at future $\gamma \gamma$ colliders.
\end{abstract}

\section{ Introduction}

Optics of high-energy $\gamma$-quanta ($\hbar \omega
\stackrel{>}{\sim}$ 100 GeV) in matter is determined mainly by
process of the $e^+e^-$ pair production. Traversing a crystal,
such $\gamma$-quanta can essentially vary their polarization
accompanying by considerable losses in intensity. These problems
including applications in high-energy physics are considered in a
number of papers (see, for example, \cite{books}, \cite{MMF} and
literature therein).

It is well known that a region with an electromagnetic field can
also be regarded as anisortopic medium (see \cite{BLP} \S
129-130). A possibility to use a bunch of laser photons as a
``crystal" is pointed out in Ref. \cite{MMF}, but the concrete
calculations of this possibility are not given.

In the present paper we study in detail properties of such a
``crystal"  considering head-on collision of hard $\gamma$-quanta
with a bunch of polarized laser photons. For the energy of
$\gamma$-quanta below the threshold of the $e^+e^-$ pair
production
\begin{equation}
\hbar \omega < \hbar \omega_{{\rm th}} =
{m_e^2 c^4\over \hbar \omega_L} = {260 \; {\rm GeV}\over (\hbar
\omega_L/ {\rm eV})}
\label{1}
\end{equation}
(here $\hbar \omega_L$ is the laser photon energy and $m_e$
is the electron mass) the main interaction is the elastic
light-light scattering $\gamma \gamma_L \to \gamma \gamma_L$.
Cross section of this process $\stackrel{<}{\sim} \alpha^2\,
r^2_e$ is approximately by 5 order of magnitude less then
a typical cross section for the pair production  $\sim \pi r^2_e$,
where $r_e=e^2/(m_ec^2)$ is the classical electron radius.
Therefore, the laser bunch is practically transparent for such
$\gamma$-quanta. On the other hand, the variation in polarization
for the $\gamma$-quantum traversing the bunch is determined by
the interference of the incoming wave and the wave scattered at
zero angle. In other words, for such a variation it is
responsible not the cross section (which is proportional to
{\bf square of the light-light scattering amplitude} of the order
of $\alpha^4$), but {\bf the scattering amplitude itself} $\sim
\alpha^2$, where $\alpha = e^2/(\hbar c)=1/137$. As a result, in
this case the essential variation in the $\gamma$-quantum
polarization can occur practically without loss in intensity of
$\gamma$-quanta.

This effect can be interesting for the following reasons:

(i) It can be used to modify the polarization of hard
$\gamma$-quanta without loss in their intensity. In particular,
with its help it is possible to transform the circular
polarization into the linear one or the linear polarization into
the circular one, and it is possible to rotate the direction of
the linear polarization.

(ii) The experimental observation of variation in the
$\gamma$-quantum polarization at passage through the bunch of
polarized laser photons below the threshold of the $e^+e^-$ pair
production will be indirect observation the process of the
elastic light-light scattering. The conditions close to those
necessary for observation of such effect is realized now at SLAC
in E-144 experiment \cite{E144}.

(iii) The discussed problem is also actual in connection with
projects of $\gamma \gamma$ colliders which under development now
(see Refs. \cite{gg}, \cite{Berkeley}, \cite{ZDP}). In these
projects it is suggested to obtain the required high-energy
$\gamma$-quanta by backward Compton scattering of laser light on
the electron beam of a linear collider. The planned almost the
whole conversion $e \to \gamma$ will take place under condition
that an electron travels in a laser ``target" an optical
thickness $t$ of the order of one:
$$
t \sim \pi\, r^2_e\, n_L \, l_L \; \sim \; 1,
$$
where $n_L$ is the concentration of the laser photons in the
bunch and  $l_L$ is its length. The $\gamma$-quantum obtained
inside the bunch will travel further through the same bunch
varying its polarization. We will show that this variation is
determined approximately by the same parameter $t$ and it may be
quite essential. Therefore, such a variation in polarization
should be, generally speaking, taken into account at simulations
of the $e \to \gamma$ conversion process performed just now for
such colliders (see, for example, \cite{ZDP}).  Besides, if one
adds in the scheme of the  $\gamma\gamma$ collider the laser
flash of the defined polarization, one can exert control over the
$\gamma$-quantum polarization.

\section{Equations for Stokes parameters of the
$\gamma$-quantum}

Let us consider the head-on collision of $\gamma$-quanta with the
bunch of laser photons. We choose the $z$ axis along the momenta
of $\gamma$-quanta. The polarization state of $\gamma$-quantum is
described by Stokes parameters $\xi_{1,2,3}$, among them $\xi_2$
is the degree of the circular polarization (which is equal to the
mean $\gamma$-quantum helicity) and $\sqrt{\xi_1^2 +\xi_3^2}$ is
the degree of the linear polarization. In the helicity basis
($\lambda,\; \lambda' = \pm 1$) the density matrix of
$\gamma$-quantum has the form (see, for example, Ref. \cite{BLP}
\S 8):
\begin{eqnarray}
\rho _{\lambda \lambda '}^\gamma = {1\over 2}
\left(
\begin{array}{cc}
1+\xi_2           &  -\xi_3 +i \xi_1 \\
-\xi_3 - i \xi_1  &  1-\xi_2
\end{array}
\right) \,.
\label{2}
\end{eqnarray}
For the laser photon such a matrix is
described by the following parameters: the degree of the circular
polarization $P_c$, the degree of the linear polarization $P_l$
and the direction of the linear polarization. Let us choose the
$x$ axis along this direction \footnote{We restrict ourselves to
the case when the direction of the linear polarization is
constant inside the laser flash. More general case corresponds
only to a little more cumbersome expressions.}, then
\begin{eqnarray}
\rho _{\lambda \lambda '}^L = {1\over 2}
\left(
\begin{array}{cc}
1+P_c & -P_l \\
-P_l  & 1-P_c
\end{array}
\right) \,.
\label{3}
\end{eqnarray}
We will also use a compact expression describing the polarization
of both photons
\begin{equation}
\rho_{\Lambda \Lambda'} = \rho^\gamma_{\lambda_1\lambda_1'}\;
\rho^L_{\lambda_2 \lambda_2'}\,.
\label{4}
\end{equation}

We will obtain further equations for Stokes parameters $\xi_i$ of
$\gamma$-quantum traversing a laser bunch. As is well known the
variations in intensity and polarization of the wave passing
through a medium are due to interference it with the wave
scattered at zero angle. Let the incoming wave has the form
$$
A_{\Lambda}\,{\rm e}^{ikz}\,.
$$
Here the amplitude $A_{\Lambda}$ describes the polarization state
of the $\gamma$-quantum and the laser photon, the wave vector
$k=\omega /c$ (the frequency of laser photon $\omega_L \ll
\omega$). When the wave passes through a ``target" layer of a
thickness $dz$ it is appeared the forward scattered wave
\begin{equation}
f_{{\Lambda \Lambda'}} \,A_{\Lambda'}\,2n_L \, dz\, \int { {\rm
e}^{ikr} \over r}\, dx\, dy =
{2\pi \, i \over k}\, f_{\Lambda \Lambda'} \,A_{\Lambda'}\,2n_L
\, dz\, {\rm e}^{ikz} \, =
dA_{\Lambda}\,{\rm e}^{ikz}\, ,
\label{5}
\end{equation}
where $f_{{\Lambda \Lambda'}}$ is the forward amplitude for the
process of elastic scattering light by light. The factor 2 in
front of $n_L$ is due to relative motion of the $\gamma$-quanta
and the ``target".

The matrix $\rho_{{\Lambda \Lambda'}}$ is expressed trough the
product of $A_{\Lambda}$:
\begin{equation}
\rho_{\Lambda \Lambda'} = {\langle A_{\Lambda} A^*_{\Lambda'}
\rangle \over N } \,, \;\;\;\;\;
N=     \langle A_\Lambda A^*_{\Lambda} \rangle \, ,
\label{6}
\end{equation}
where  $\langle ... \rangle$ denotes a statistical averaging. The
quantity $N$ is proportional to the $\gamma$-quantum intensity
$J$. When the wave passes through the layer of a thickness $dz$
its relative variation in intensity is equal to
\begin{equation}
{dJ\over J} ={dN\over N}= {2\over N}\, {\rm Re}\, \langle
dA_\Lambda A^*_{\Lambda} \rangle \, = -{4\pi \over k}\, {\rm
Im}\,(f_{\Lambda \Lambda'} \, \rho_{\Lambda' \Lambda}) \, 2n_L\,
dz \,.
\label{7}
\end{equation}
If we introduce the total cross section for the light-light
scattering
\begin{equation}
\sigma_{\gamma \gamma} = {4\pi \over k}\, {\rm Im}\, (f_{\Lambda
\Lambda'} \, \rho_{\Lambda' \Lambda}) \, ,
\label{8}
\end{equation}
then the Eq. (\ref{7}) can be presented in the form
\begin{equation}
dJ = -\,\sigma_{\gamma \gamma}\,2n_L dz \, J\,.
\label{9}
\end{equation}
Analogously,
\begin{equation}
d\rho_{\Lambda \Lambda'} = d{\langle A_{\Lambda} A^*_{\Lambda'}
\rangle \over N } \, = {2\pi\,i \over k}\,
(f_{\Lambda \Lambda''}\rho_{\Lambda'' \Lambda'} -
f^*_{\Lambda' \Lambda''}\rho_{\Lambda \Lambda''}) \, 2n_L dz-
\rho_{\Lambda \Lambda'}\, {dN\over N}\,.
\label{10}
\end{equation}

Instead of the scattering amplitudes $f_{\Lambda \Lambda'}$ it is
convenient to use the invariant scattering amplitudes
$M_{\lambda_1 \lambda_2\, \lambda'_1 \lambda'_2}$ defined in Ref.
\cite{BLP} \S 127
\begin{equation}
f_{\Lambda \Lambda'} \equiv
f_{\lambda_1 \lambda_2\, \lambda'_1 \lambda'_2}=
{k\over 4\pi}\, {(\hbar c)^2 \over s}\,
M_{\lambda_1 \lambda_2\, \lambda'_1 \lambda'_2}\, , \;\;\;
\; s=4\,\hbar \omega\, \hbar \omega_L\,,
\label{11}
\end{equation}
where quantity $s$ is the square of the total energy of the
$\gamma$-quantum and the laser photon in their centre-of-mass
system. Among the five independent helicity amplitude only three
ones are not equal to zero for the forward scattering, namely
$$
M_{++++}=M_{----}\,,\;\; M_{+-+-}=M_{-+-+}\,, \;\; M_{++--}
=M_{--++}\,.
$$
We will use further the following real quantities $R$ and $I$
proportional correspondingly to the real and imaginary parts of
the scattering amplitudes divided by $\alpha^2 s$:
$$
I_{np}={(\hbar c)^2\over s}\,{{\rm Im}\, (M_{++++}\,+\, M_{+-+-})\over
2\pi r^2_e },
$$
\begin{equation}
R_c +iI_c={(\hbar c)^2\over s}\,{M_{++++}\,-\, M_{+-+-}\over
2\pi r^2_e }, \;\;\;\;
R_l +iI_l={(\hbar c)^2\over s}\,{M_{++--}\over
2\pi r^2_e }\, .
\label{12}
\end{equation}

By substituting Eqs. (\ref{4}), (\ref{11}), (\ref{12}) into Eqs.
(\ref{7}), (\ref{8}), (\ref{10}) we shall obtain the expression
for the cross section
\begin{equation}
\sigma_{\gamma \gamma} = \pi r^2_e\, (I_{np} +\xi_2 P_c \, I_c +
\xi_3 P_l \, I_l)
\label{13}
\end{equation}
and equations for Stokes parameters. To write down these
equations it is convenient to introduce the quantity
\begin{equation}
dt = 2\pi r^2_e \, n_L dz
\label{14}
\end{equation}
which we will call the reduced optical thickness of the layer
$dz$. Then
$$
{d\xi_1\over dt} = (-R_c \xi_3 +I_c \xi_1 \xi_2) \,P_c \,+\,
(R_l \xi_2 +I_l \xi_1 \xi_3)\,P_l\, ,
$$
\begin{equation}
{d\xi_2\over dt} = - I_c(1- \xi_2^2)\, P_c \,+\,
(-R_l \xi_1 +I_l \xi_2 \xi_3)\,P_l\, ,
\label{15}
\end{equation}
$$
{d\xi_3\over dt} = (R_c \xi_1 +I_c \xi_2 \xi_3) \,P_c \,-\,
I_l (1- \xi_3^2)\,P_l\, .
$$

Integrating these equations one can obtain the dependence of
Stokes parameters on the optical thickness $t$. After that the
dependencies of cross section (\ref{13}) and then the
intensity (\ref{9}) on $t$ are determined.

The physical meaning of different items in the cross section
(\ref{13}) can be easily established if one considers the
collisions of photons in pure quantum states. Let $\sigma_0$ and
$\sigma_2$ denote the cross sections for collisions of photons
with the total angular momentum $J_z = \xi_2 -P_c$ equals 0 and
2, and $\sigma_\parallel$ and $\sigma_\perp$ denote the cross
sections for collisions of photons with parallel ($\xi_3 =P_l
=1$) and orthogonal ($\xi_3 = -P_l =1$) linear polarizations.
Then the quantity $I_{np}$ corresponds to the cross section for
nonpolarized photons
\begin{equation}
\sigma_{np} = \pi r^2_e\,I_{np} = {1\over 2} (\sigma_0 +\sigma_2)
= {1\over 2} (\sigma_\parallel +\sigma_\perp )\, ,
\label{16}
\end{equation}
and quantities $I_c$ and $I_l$ correspond to asymmetries for the
circular and linear polarization respectively
\begin{equation}
\pi r^2_e \, I_c = {1\over 2} (\sigma_0 -\sigma_2) \,, \;\;\;
\pi r^2_e \, I_l ={1\over 2} (\sigma_\parallel -\sigma_\perp )\,.
\label{17}
\end{equation}

The forward scattering amplitudes (and, therefore, quantities $R$
and $I$) depend on the single variable
$$
r ={s\over 4m_e^2 c^4} ={\omega \over \omega_{{\rm th}}}\,.
$$
Using for amplitudes $M_{\lambda_1 \lambda_2\, \lambda'_1
\lambda'_2}$ formulas from Refs.  \cite{Tol} and \cite{BLP} \S
127 we obtain the following expressions for functions (\ref{12}):
$$
I_{np}= 0\;\; {\rm at}\; r<1\,; \;\;
I_{np}={1\over r} \left[2 \left( 1+ {1\over r}
-{1\over 2r^2} \right) {\rm cosh^{-1}}\sqrt{r} - \left( 1+ {1\over r}
\right) \sqrt{ 1- {1\over r}} \right] \;\;   {\rm at} \; r > 1
\,,
$$
\begin{equation}
R_c +i I_c = {2\over \pi r} (-3B_- +T_- ) , \;\;
R_l +i I_l = {1\over \pi r} \left( 1 + {1\over r}B_- +
{1\over 2r^2} T_+ \right) \,,
\label{18}
\end{equation}
where
$$
B_-\, =\,
\left\{
\begin{array}{rl}
{\displaystyle
\sqrt{{1\over r}-1} \sin^{-1}{\sqrt{r}}- \sqrt{{1\over r}+ 1}\,
{\rm sinh^{-1}}{\sqrt{r}}
}
& {\rm at} \; r < 1 \\
{\displaystyle
\sqrt{1-{1\over r}} \,{\rm cosh^{-1}}{\sqrt{r}}- \sqrt{{1\over
r}+ 1}\, {\rm sinh^{-1}}{\sqrt{r}} -i {\pi\over 2} \sqrt{1-{1\over r}}
}
& {\rm at} \; r > 1 ,
\end{array}
\right.
$$
$$
T_{\pm}\, =\,
\left\{
\begin{array}{rl}
{\displaystyle
-(\sin^{-1}{\sqrt{r}})^2\, \pm \, ({\rm sinh^{-1}}{\sqrt{r}})^2
}
& {\rm at} \; r < 1 \\
{\displaystyle
-{\pi^2\over 4} \,+ \,({\rm cosh^{-1}}{\sqrt{r}})^2 \,\pm \,
({\rm sinh^{-1}}{\sqrt{r}})^2 -i \pi \,{\rm cosh^{-1}}{\sqrt{r}}
}
& {\rm at} \; r > 1 \,.
\end{array}
\right.
$$
Functions $R$ and $I$ are presented in Figs. 1 and 2. Note that
extremums of $R_c$ and $R_l$ are at the threshold of the pair
production:
\begin{equation}
R_c=0.315, \;\;\; R_l=-0.348 \;\;\; {\rm at} \; s = 4m_e^2c^4\,.
\label{19}
\end{equation}
In the region below the threshold these functions decrease
very rapidly with decreasing of $s$:
\begin{equation}
R_c={64\over 315 \pi}\, r^2, \;\;\;\;
R_l=-{4\over 15 \pi}\,r \;\;\;\; {\rm at}\; \; \; r\ll 1.
\label{20}
\end{equation}

\section{A laser bunch as a transparent anisotropic medium}

A laser bunch becomes transparent for $\gamma$-quanta with the
energy below the threshold of the pair production $\omega \, <
\, \omega_{{\rm th}}$ (see Eq. (\ref{1})):
$$
I_{np}=I_c=I_l =\sigma_{\gamma \gamma} =0
\;\;\;{\rm at} \;\;\;s < 4m^2_e c^4.
$$

If the laser photons are linearly polarized ($P_l \neq 0, \; P_c=
0$) the solution of Eqs. (\ref{15}) has the form
\begin{equation}
\xi_1 =\; \xi_1^0 \cos{\varphi_l}+\xi_2^0 \sin{\varphi_l}\,,\;\;
\xi_2 = -\xi_1^0 \sin{\varphi_l}+\xi_2^0 \cos{\varphi_l}\,, \;\;
\xi_3 =\; \xi_3^0 \,,
\label{21}
\end{equation}
where the phase $\varphi_l = P_l\,R_l\,t$ and $\xi_i^0$ are
the initial Stokes parameters. It is seen from this solution that
in this case the laser bunch is an anisotropic medium with
different refraction indices $n_x$ and $n_y$ along the $x$ and
$y$ axes:
\begin{equation}
n_x-n_y = {c\over \omega}\,  2\pi r_e^2\, n_L \,P_l\, R_l.
\label{22}
\end{equation}
Such a medium transforms the circular polarization of
$\gamma$-quanta into the linear one and vice versa. If, for
example, the initial $\gamma$-quantum is circularly polarized,
$\xi_2^0 \neq 0,\, \xi_1^0 = \xi_3^0 =0$, its polarization
transforms to the linear one, $\xi_1 = - \xi_2^0\,, \;
\xi_2=\xi_3 =0$, when the phase $\varphi_l$ becomes equal to $-
\pi /2$.

If the laser photons are circularly polarized ($P_c \neq 0,
\;P_l=0$), the solution of Eqs. (\ref{15}) has another form
\begin{equation}
\xi_1 = \xi_1^0 \cos{\varphi_á}-\xi_3^0 \sin{\varphi_á}\,,\;\;
\xi_2 = \xi_2^0 \,,\;\;
\xi_3 =\xi_1^0 \sin{\varphi_á}+\xi_3^0 \cos{\varphi_á}\,,
\label{23}
\end{equation}
where the phase $\varphi_á= P_c\,R_c\, t$. From this solution it
is seen that in this case the laser bunch is an gyrotropic
medium with different refraction indices $n_+$ and $n_-$ for the
right and left polarized $\gamma$-quanta, and
$$
n_+ -n_- = {c\over \omega}\,  2\pi r_e^2\, n_L \,P_c\, R_c.
$$
Such a medium rotates the direction of the $\gamma$-quantum linear
polarization on the angle $(-\,\varphi_á /2)$.

In a general case
\begin{equation}
\xi_1 =\; A  \cos{(\varphi + \varphi_0)}\,,\;\;\;
\xi_2 = - {P_l\, R_l\over R}\,A  \sin{(\varphi + \varphi_0)}
\,+\, {P_c R_c\over R}\, B\,,
\label{24}
\end{equation}
$$
\xi_3 = \; {P_c\, R_c\over R}\,A  \sin{(\varphi + \varphi_0)}
\,+\, {P_l R_l\over R}\, B\,,\;\;\;
R =\sqrt{(P_c\,R_c)^2 +(P_l\,R_l)^2}\,,
$$
where the phase $\varphi = R t$ and constants $A$, $B$ and
$\varphi_0$ are determined by the initial conditions. Note that
the total degree of the $\gamma$-quantum polarization is not
changed:
$$
\sqrt{\xi_1^2 +\xi_2^2 +\xi_3^2} = \sqrt{A^2+B^2} =
{\rm const}\,.
$$

\section{Variation in polarization above the threshold of the pair
production}

Above the threshold of the $e^+e^-$ pair production ($\omega \, >
\, \omega_{{\rm th}}$ ) the variation in the $\gamma$-quantum
polarization is accompanied by a reduction in their intensity in
accordance with Eq. (\ref{9}). In this case the total
$\gamma$-quantum degree of polarization are not conserved. We
give the solutions of Eqs. (\ref{15}) for two particular cases
discussed in the above section.

If the laser photons are linearly polarized ($P_l \neq 0, \; P_c=
0$) then
\begin{equation}
\xi_1 =\; (\xi_1^0 \cos{\varphi_l}+\xi_2^0
\sin{\varphi_l})\,/D_l\,,\;\;
\xi_2 = (-\xi_1^0 \sin{\varphi_l}+\xi_2^0
\cos{\varphi_l})\,/D_l\,,
\label{25}
\end{equation}
$$
\xi_3 =\; (\xi_3^0 {\rm ch}\tau_l - {\rm sh}\tau_l )\,/D_l\,,\;\;
D_l= {\rm ch}\tau_l - \xi_3^0 {\rm sh}\tau_l \,,\;\;\;\tau_l =
P_l\,I_l\,t\,.
$$
The similar case (in connection with the problem of passage of
$\gamma$-quanta through a monocrystal) was considered in detail
in Ref. \cite{MMF}. Note that $\xi_3 \to 1$ at $\tau_l \to
- \infty$.

If the laser photons are circularly polarized ($P_c \neq 0,
\;P_l=0$) then
\begin{equation}
\xi_1 = (\xi_1^0 \cos{\varphi_á}-\xi_3^0
\sin{\varphi_á})\,/D_c\,,\;\;
\xi_2 = (\xi_2^0 {\rm ch}\tau_c - {\rm sh}\tau_c )\,/D_c\,,
\label{26}
\end{equation}
$$
\xi_3 =(\xi_1^0 \sin{\varphi_á}+\xi_3^0
\cos{\varphi_á})\,/D_c\,,\;\;
D_c= {\rm ch}\tau_c - \xi_2^0 {\rm sh}\tau_c \,,
\;\;\;\tau_c = P_c\,I_c\,t\,.
$$
In this case $\xi_2 \to \pm 1$ at $\tau_á \to \mp \infty$.

\section{Discussion}

{\bf 1.} As a result, we have shown that below the threshold of
the $e^+e^-$ pair production the laser bunch is similar to the
transparent anisotropic medium. In particular, the linearly
polarized bunch corresponds to the uniaxial crystal and the
circularly polarized bunch corresponds to the gyrotropic
medium. Let us illustrate the magnitude of the discussed effects
using as an example the parameters of the laser bunch given in
Ref. \cite{ZDP} (they are close to the parameters which are
realized in the experiment E-144 at SLAC \cite{E144}):  $\hbar
\omega_L =1.18$ eV, the energy of the laser flash is 1 J, the
laser bunch length is 1.8 ps, and the peak intensity is about
10$^{18}$ W/cm$^2$. The reduced optical thickness (\ref{14}) for
this flash is equal to $t = 1.4$. As is seen from Fig. 1
phases $\varphi_l = P_l\,R_l\, t$ and $\varphi_á = P_á\,R_á\, t$
which determine the magnitude of the effect can reach values
$\approx 0.3 t \sim 1$. According to Eqs. (\ref{21}) and
(\ref{23}) it means that the variation in the $\gamma$-quantum
polarization may be very large. It is also seen from Fig. 1 that
the effect depends strongly on the energy and it becomes very
small at $\omega \ll \omega_{{\rm th}}$.

{\bf 2.} With the growth of the intensity of laser flash it is
necessary to take into account the effects of intense
electromagnetic fields (see Ref. \cite{NQED} and literature
therein) which we neglect in the present paper.

{\bf 3.} In the scheme of the $e \to \gamma$ conversion adopted
for the $\gamma \gamma$ colliders, $\gamma$-quanta are produced
inside the laser bunch. When such $\gamma$-quanta travel further
in the laser bunch they can essentially vary their polarization.
It should be noted, however, that for the optimal conversion the
laser photons and the hardiest $\gamma$-quanta are circularly
polarized \cite{gg}, \cite{ZDP}. These $\gamma$-quanta conserve
their polarization on the rest way through the bunch. But the
$\gamma$-quanta with a lower energy have a fraction of the
linear polarization, and the rotation of the direction of this
linear polarization should be, generally speaking, taken into
account.

{\bf 4.} We apply the same method to calculate the variation in
the polarization of electrons traversing through a bunch of
polarized laser photons. The corresponding results will be given
in a separate paper.

We are very grateful to I. Ginzburg and A. Onuchin for useful
discussions.

\vspace{1cm}

\centerline{{\bf Figure captions}}

\vspace{0.3cm}

{\bf Fig. 1.} The real parts of the scattering amplitudes for the
light-light scattering at zero angle (see Eqs. (\ref{12})
(\ref{18})) in dependence on the parameter $s/(m_e^2 c^4)=
4\omega /\omega_{{\rm th}}$.

\vspace{0.3cm}

{\bf Fig. 2.} The same for the imaginary parts of amplitudes. The
quantity $\pi\,r_e^2\,I_{np}$ is equal to the cross section of
the $\gamma \gamma_{L} \to e^+ e^-$ process for nonpolarized
particles.

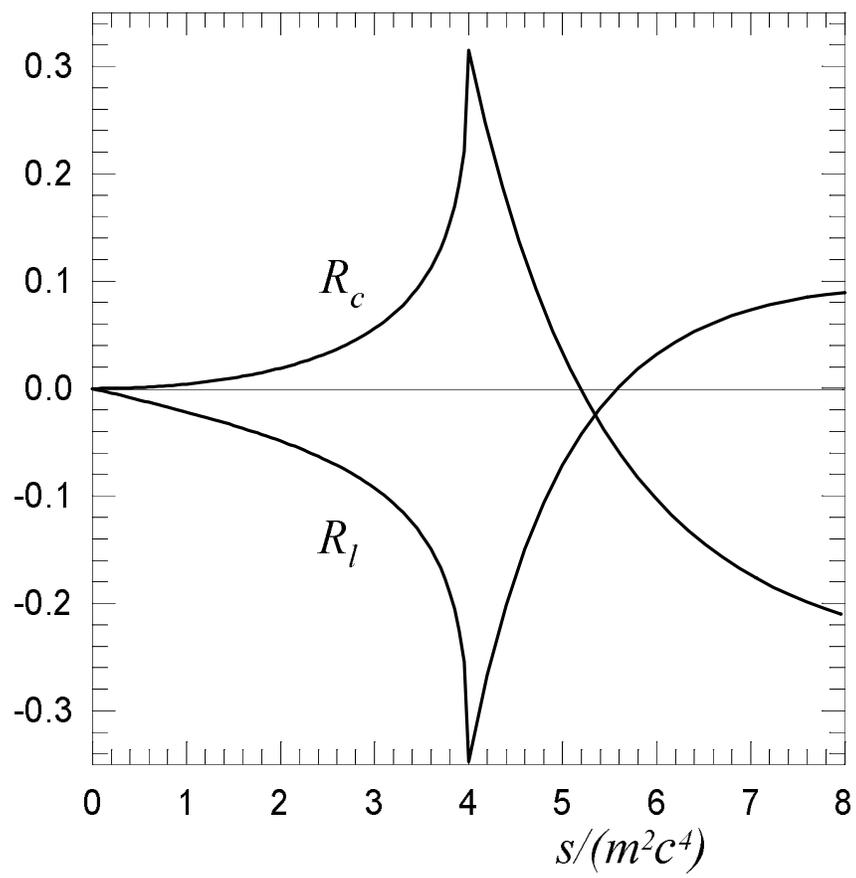
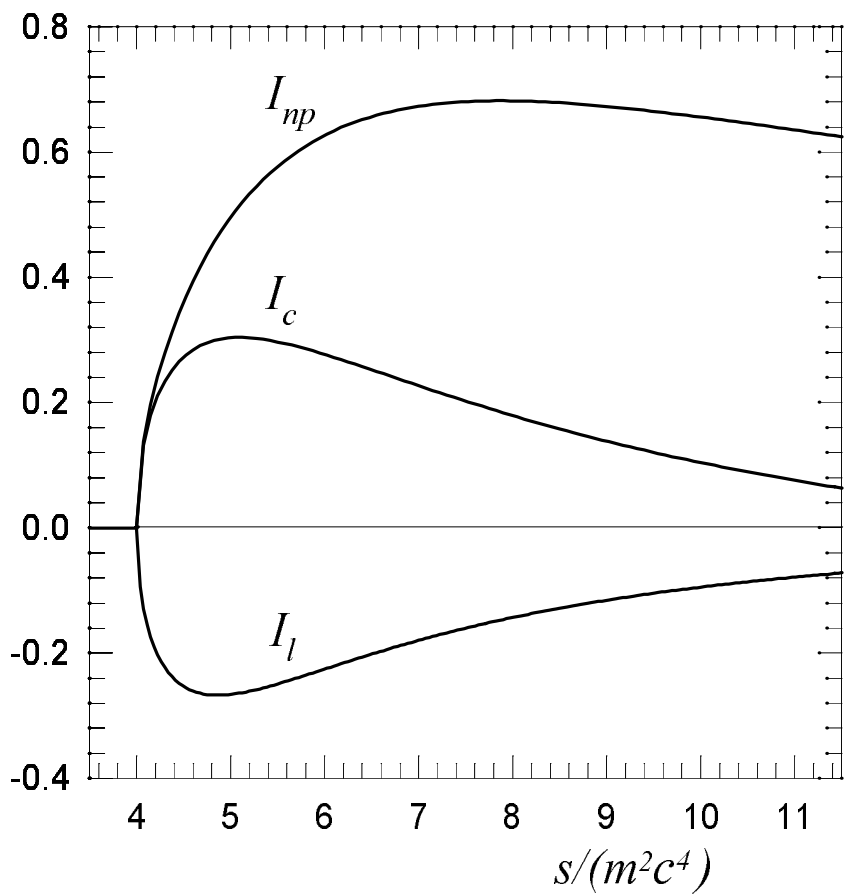
\end{document}